\documentstyle[11pt,paspconf,psfig]{article}
\def\cor{\mathrel{\mathchoice {\hbox{$\widehat=$}}{\hbox{$\widehat=$}}
{\hbox{$\scriptstyle\hat=$}}
{\hbox{$\scriptscriptstyle\hat=$}}}}
\def\la{\mathrel{\mathchoice {\vcenter{\offinterlineskip\halign{\hfil
 $\displaystyle##$\hfil\cr<\cr\sim\cr}}}
 {\vcenter{\offinterlineskip\halign{\hfil$\textstyle##$\hfil\cr
 <\cr\sim\cr}}}
 {\vcenter{\offinterlineskip\halign{\hfil$\scriptstyle##$\hfil\cr
 <\cr\sim\cr}}}
 {\vcenter{\offinterlineskip\halign{\hfil$\scriptscriptstyle##$\hfil\cr
 <\cr\sim\cr}}}}}
\def\ga{\mathrel{\mathchoice {\vcenter{\offinterlineskip\halign{\hfil
 $\displaystyle##$\hfil\cr>\cr\sim\cr}}}
 {\vcenter{\offinterlineskip\halign{\hfil$\textstyle##$\hfil\cr
 >\cr\sim\cr}}}
 {\vcenter{\offinterlineskip\halign{\hfil$\scriptstyle##$\hfil\cr
 >\cr\sim\cr}}}
 {\vcenter{\offinterlineskip\halign{\hfil$\scriptscriptstyle##$\hfil\cr
 >\cr\sim\cr}}}}}

\newcommand{\Lsun}{\,\mbox{L}_{\odot}}

\newcommand{\etal}{{\it et al.\/}~}

\newcommand{\ie}{{\it i.e.},\ }
\newcommand{\eg}{{\it e.g.},\ }

\newcommand{\Mzon}{M$_{\odot}$}
\newcommand{\Lzon}{L$_{\odot}$}

\newcommand{\kms}{\mbox{km s$^{-1}$}}

\newcommand{\Mpc}{Mpc$^{-1}$}

\newcommand{\HI}{{\sc H$\,$i}}
\newcommand{\HII}{{\sc H$\,$ii}}

\begin{document}
 
\title{Properties of Dust in Giant Elliptical Galaxies: \\ The
r$\hat{\mbox{o}}$le of the Environment} 

\author{Paul Goudfrooij\,\altaffilmark{*}}

\affil{Space Telescope Science Institute, 3700 San Martin Drive, 
Baltimore, \\ MD 21218, U.S.A.}

\altaffiltext{1}{Affiliated to the Astrophysics Division, Space
Science Department, European Space Agency}

\begin{abstract}
The properties of dust in giant elliptical
galaxies are reviewed, with particular emphasis on the influence of the 
environment.  

After normalizing by the optical luminosities, a strong 
anticorrelation between the masses of dust and hot
gas in X-ray bright ellipticals is found.  
Furthermore, large-scale, regularly-shaped dust lanes (which are
symmetric with respect to the galaxy nucleus) 
are only found to be present in ellipticals with the
lowest ratios of the mass of hot gas to the blue luminosity
($M_{{\rm hot}}/L_B \la 0.04$ in solar units). 
This can be explained by the short time scale for the destruction of
dust grains within the hot, X-ray-emitting gas compared to the
formation timescale of such dust lanes in early-type galaxies.  

Dust within ellipticals in ``loose'' environments (\ie in the field or in
loose groups) is typically characterized by small values of $R_V \equiv
A_V/E_{B-V}$ (\ie small characteristic grain sizes), and distributed in
dust lanes with a smooth, relaxed morphology. On the other hand, dust in
ellipticals that are in or near the center of dense groups or clusters is
typically much more irregularly distributed, and characterized by $R_V$
values that are close to the Galactic one.  

I predict that ellipticals containing ``relaxed'' dust lanes with typical
dust masses of $10^6 - 10^7$ M$_{\odot}$ do not contain hot, X-ray-emitting
gas.  

\end{abstract}
 

\addtocounter{footnote}{1}
 
\section{Introduction:\ The Multi-Phase ISM of Elliptical Galaxies} 

It has become evident in recent years that elliptical galaxies are far from
being the simple, (violently) relaxed, isothermal, purely stellar systems
anticipated by the traditional picture developed by Hubble. 
%
Recent deep surveys across the
electromagnetic spectrum have shown that ellipticals contain a complex,
multi-phase interstellar medium (ISM).  In fact, all ISM components known to
exist in spiral galaxies are 
now accessible in elliptical galaxies as well, although in rather different
proportions.  The main difference is that the dominant (in mass) gaseous
component in spirals is ``cool'', in the form of neutral gas (\HI, H$_2$),
whereas in ellipticals it is ``hot'' ($\sim 10^7$ K), radiating at X-ray
wavelengths. The typical mass of this hot gas has been found to be of order
$10^9 - 10^{11}$ \Mzon\ (for H$_0$ = 50 \kms\ \Mpc), which is similar
to the expected amount from stellar mass loss accumulated over 
the lifetime of luminous ellipticals (\eg Forman \etal 1985;
Loewenstein 1998). Hence, it is now commonly believed that
the hot gas is indeed originated ---and constantly replenished--- by mass
loss of stars within the ellipticals which is thermalized in the
gravitational field of the galaxies (\ie with temperature $T_{\sigma} =
\mu m_p\sigma_*^2/k$, where $\mu$ is the mean atomic weight, $m_p$ is the
proton mass, and $\sigma_*$ is the stellar velocity dispersion). 

It should be noted that evidence for the existence of this hot
component of the ISM is currently only substantial for the most luminous
ellipticals ($L_B \ga 5 \; 10^{10}\:$\Lzon, see \eg Kim \etal 1992), as well
as for ellipticals that are in the centers of 
groups (Mulchaey \& Zabludoff 1998). 
It thus seems that only those ellipticals that are ``privileged'' to reside
within a deep enough potential well are able to retain the
material lost by stars and suppress the supernova-driven wind 
proposed by Mathews \& Baker (1971). In this scenario, smaller ellipticals
with too shallow potential wells may not see the bulk of their internally
produced ISM ever again, donating it to the intracluster or
intragroup medium. 

Along with the ``hot'' ISM, the cooler ISM components exist in
ellipticals as well:\  
``warm'' ionized gas (Goudfrooij 1998; Macchetto \etal 1996), dust (Knapp
\etal 1989; Goudfrooij \& de Jong 1995), and ``cold'' CO and H\,{\sc i}
(Lees \etal 1991; Wiklind \etal 1995; Oosterloo \etal 
1998). However, the measured amounts of dust, cold gas and warm gas are 
typically small, and do not correlate with the stellar luminosity of 
ellipticals (as opposed to the case among spirals, cf.\ Lees \etal
1991). While it is conceivable that some part of the observed dust and 
gas originates from stellar mass loss, this observational result indicates
that most of it has an external origin (\eg accreted during an interaction
with a smaller, gas-rich galaxy). See also Forbes (1991) and Goudfrooij \& de
Jong (1995). 

Typical  properties of the ISM in ellipticals are listed below in Table 1
(for comprehensive reviews, see \eg Goudfrooij 1997, 1998; Knapp 1998). 

\vspace*{-1.5ex}
\begin{table}[h]
\caption{The Multi-Phase ISM of Elliptical Galaxies}
{\footnotesize 
\begin{tabular*}{10.9cm}{@{\extracolsep{\fill}}lrrcc@{}}
\multicolumn{5}{c}{~} \\ [-1.8ex]   \hline \hline 
\multicolumn{5}{c}{~} \\ [-1.8ex]
ISM Phase & \multicolumn{1}{c}{Temp.} & \multicolumn{1}{c}{$f^a$} & 
  \multicolumn{1}{c}{Mass}  & Probable Origin \\ [0.5ex] \hline
\multicolumn{5}{c}{~} \\ [-1.8ex]
Molecular & $\la$\,10 K    & $\sim$\,8\%  & $\la\,10^7$ \Mzon   & External \\
Dust 	  & $\sim$\,15$-$100 K & $\ga$\,45\%  & $10^4 - 10^7$ \Mzon & 
	External \\
\HI 	  & $\sim$\,100 K  & $\sim$\,10\% & $10^8 - 10^9$ \Mzon & External \\
\HII	  & $\sim\,10^4$ K & $\sim$\,70\% & $10^3 - 10^5$ \Mzon & External \\
Hot	  & $\sim\,10^7$ K & $\sim$\,70\% & $10^9 - 10^{11}$ \Mzon & 
  Internal \\ 
\multicolumn{5}{c}{~~} \\ [-1.8ex] \hline
\multicolumn{5}{c}{~~} \\ [-1.2ex]
\multicolumn{5}{l}{$^a$ $f \cor$ Detection fraction} \\
\end{tabular*}
}
\end{table}

\vspace*{-2.5ex}
\section{Properties of Dust within Ellipticals in Different Environments}

Dust features have been found to exist in about half of all nearby
ellipticals. 
Dust lanes or patches are known to exist in small ellipticals as
well as in giant ellipticals, whereas the latter often reside in
environments of high galaxy density and/or massive halos of hot gas. 
This leads me to
the main issue I wish to  discuss in this contribution: Do the properties of
dust (\eg grain size, dust content, morphology) vary among ellipticals due
to differences in the environment\,?  

Most detailed studies of the properties of dust in ellipticals have been
undertaken by means of {\it optical\/} imaging, thanks to its intrinsically
high spatial resolution and the inherently smooth optical light
distributions of ellipticals which eases the comparison of extinguished
areas of the galaxies  with the appropriate unextinguished areas (e.g.,
Goudfrooij \etal 1994b). As the extinction in the optical regime ($\lambda
\sim 4000-7000$ \AA) is caused by grains with a typical size of $\sim$\,0.1
$\mu$m, it is worth taking a look at the typical grain destruction 
time scales ($\tau_{{\rm d}}$) of different mechanisms that might cause any
size change for 0.1 $\mu$m-sized dust grains within ellipticals. These 
mechanisms are summarized below: \\ [-3.5ex]
\begin{enumerate}
\item {\sf Grain-grain collisions in low-velocity} ($\la$\,50 \kms) 
 {\sf shocks:} 
  $\tau_{\rm d} \sim 1\times 10^9$ yr (\eg Jones \etal 1996). This velocity
  was selected since it is the typical maximum velocity dispersion in
  nebular emission lines within ellipticals (\eg Goudfrooij 1998).  \\ [-4ex]
\item {\sf Sputtering in supernova-driven blast waves in a two-phase medium} 
 (cold dense clouds embedded within a coronal intercloud medium of low
 density): 
 $\tau_{\rm d} \sim 3 \times 10^9$ ($L/10^{10}$\,\Lzon)$^{-1}$ yr (Draine \&
 Salpeter [1979], using the current supernova explosion rate within
 ellipticals from Turatto \etal [1996] and H$_0$ = 50 \kms\ \Mpc).  \\ [-4ex]
\item {\sf Sputtering by thermal ions:} (a) ``Warm'' 10$^4$ K gas:
 $\tau_{\rm d} 
 \sim  10^{10}$ yr (Barlow 1978); (b) ``Hot'', T\,$\sim$\,10$^7$ K gas:
 $\tau_{\rm d} \simeq 2  \times 10^5 \: (n_p/\mbox{cm}^{-3})^{-1} \,
 (a/0.1\,\mu\mbox{m})$ yr (\eg Tielens \etal
 1994). \\ [-3.5ex]
\end{enumerate}
Note that the latter destruction time scale is typically only $\la 10^7$ yr
for 0.1 $\mu$m grains (and proportionally shorter for smaller grains) within
the central few kpc of X-ray bright ellipticals where the typical proton 
density $n_p \sim 0.03\,-\,0.1$ cm$^{-3}$ (see, \eg Trinchieri
\etal 1997). Sputtering by hot ions (protons, He nuclei) is therefore the
dominant destruction agent by far for dust in ellipticals embedded in hot
gas. One 
would therefore expect the dust content of ellipticals to decrease with
increasing X-ray flux, while the dust grain size distribution in X-ray
bright ellipticals is expected to be depleted in small grains. 
On the other hand, if sputtering by hot ions is {\it not\/} the dominant
grain destruction agent (\eg in ellipticals {\it not\/} embedded in hot
gas), destruction mechanism (1) above may be dominant, which preferentially
destroys {\it large\/} grains. 

\begin{figure}[th]
\centerline{
\psfig{figure=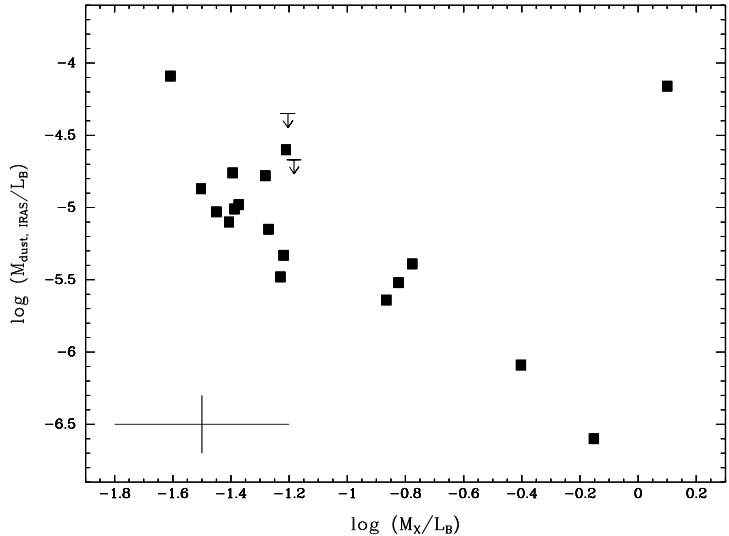,width=9.5cm,angle=-90.}
}
\vspace*{-2mm}
\caption[]{\baselineskip=0.9\normalbaselineskip
The relation of the ratio of dust mass to blue
luminosity with the ratio of the mass of hot, X-ray-emitting gas to
the blue luminosity for elliptical galaxies with $L_X/L_B \geq 2 \times
10^{30}$ erg s$^{-1}\; \mbox{L}_{\odot}^{-1}$. 
Typical errorbars are shown in the bottom left corner.}
\label{f:MXLB_MDIRLB}
\end{figure}

To address the correctness of these suggested relations, I selected from the
catalog of galaxies observed by {\sl 
EINSTEIN\/} (Fabbiano \etal 1992) the ellipticals with $L_X/L_B \geq
2 \times 10^{30}$ erg s$^{-1}\; \mbox{L}_{\odot}^{-1}$, for which Kim \etal
(1992) showed that the X-ray flux is dominated by emission from hot
gas. For those ellipticals, dust masses were derived from the {\it IRAS\/}
60 and 100 $\mu$m fluxes (Knapp \etal 1989) as described in Goudfrooij \& de
Jong (1995). Masses of hot gas were derived from the {\sl EINSTEIN\/} fluxes
according to eq.\ (9) of Canizares \etal (1987). Keep in mind that
these gas masses are uncertain by a factor of 2 -- 4 due
to the uncertainties connected with deriving density profiles from {\sl
EINSTEIN\/} data. Fig.\ \ref{f:MXLB_MDIRLB} depicts the relation of the
masses of dust and hot gas for X-ray bright ellipticals. The
masses have been divided by the optical luminosities of the galaxies to
remove the (distance)$^2$ factor. And indeed, a strong anticorrelation
between the masses of dust and hot gas is seen, as expected (cf.\ above). 

\noindent
The single data point  that deviates strongly from the general trend
in Fig.\ \ref{f:MXLB_MDIRLB} represents NGC 4696, the dominant elliptical in
the Centaurus 
cluster. This galaxy is however expected to be a special case in view of the
presence of a filamentary, obviously ``young'' dust structure in its central
region. de Jong \etal (1990) and Sparks \etal (1989) have suggested that
the dust in NGC 4696 was captured during a recent ($\sim 10^8$ yr ago) tidal
interaction with a smaller, gas-rich galaxy, while the dust can be
replenished during $\sim 10^8$ yr by evaporation of cool clouds (captured 
during the interaction) by hot  electrons within the hot gas. 

What about the dust grain size distribution\,? Goudfrooij \etal (1994b) 
measured extinction curves for dust in 10 ellipticals that display obvious
dust extinction features superposed on an otherwise smooth distribution of
light (following a de Vaucouleurs law). Interestingly, they found 
$R_V \equiv A_V/E_{B-V}$ to be very small for a number of giant ellipticals
with large-scale dust lanes having relaxed morphologies ($R_V \sim 2.1 -
2.7$, whereas $R_V = 3.1$ in the diffuse ISM in our Galaxy), which means
that the characteristic grain size for the grains causing optical extinction
is significantly smaller in those galaxies than in ours. Conversely, the
optical extinction 
curve in X-ray bright ellipticals turns out to be more ``normal'', with
$R_V$ values that are consistent with the Galactic value (Sparks \etal 1989;
Goudfrooij \& Trinchieri 1998). In Fig.\ 2, I show a comparison of the
distribution of dust extinction (through a $B\!-\!I$ image) with the
extinction curve for two 
representative cases: IC 4320, an isolated elliptical with a relaxed,
large-scale dust lane along its minor axis, and NGC 4696, the central, X-ray
bright galaxy of the Centaurus cluster (see above). 

\begin{figure}[th]
\centerline{
\psfig{figure=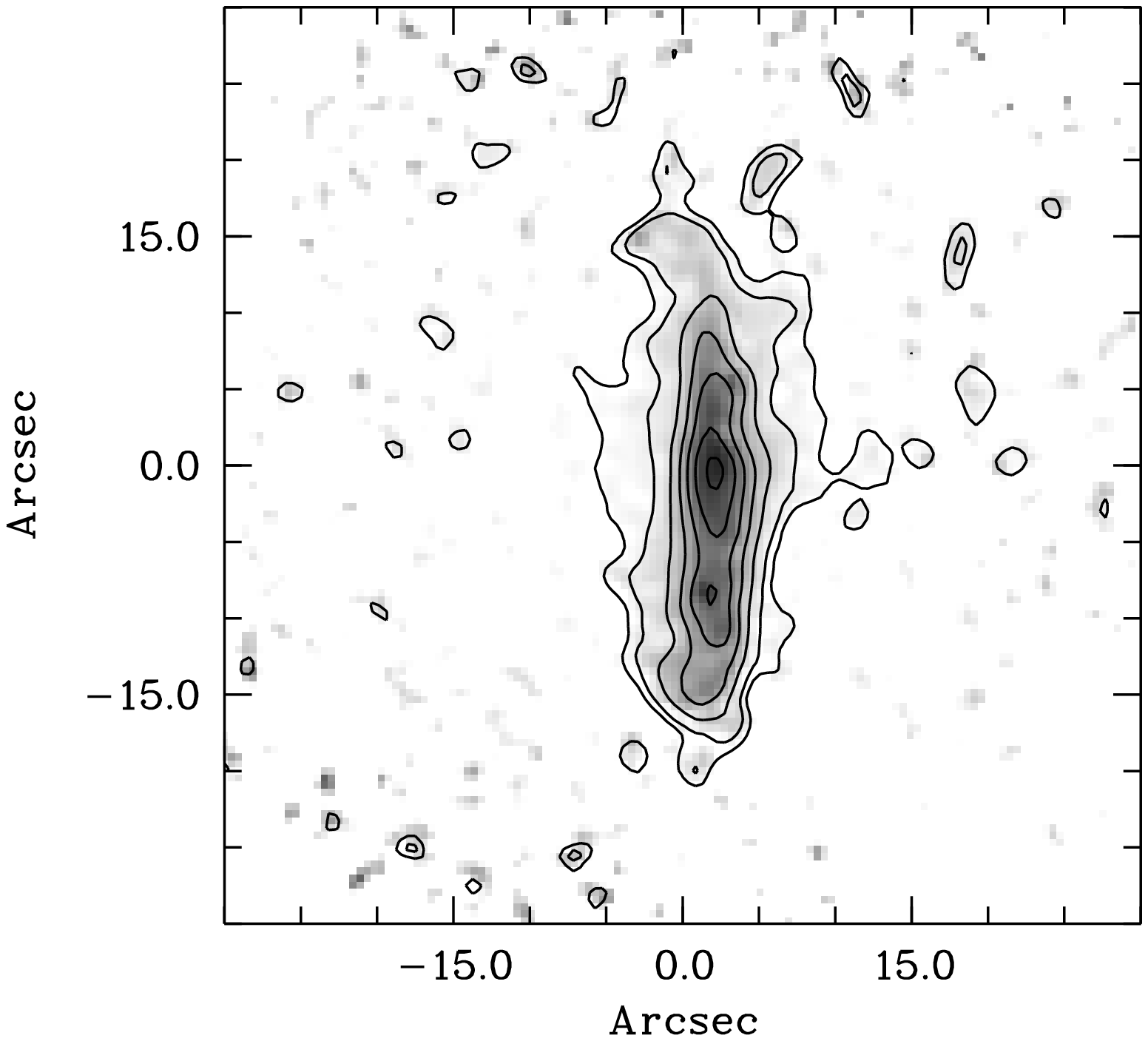,width=6.35cm}
\hfill 
\psfig{figure=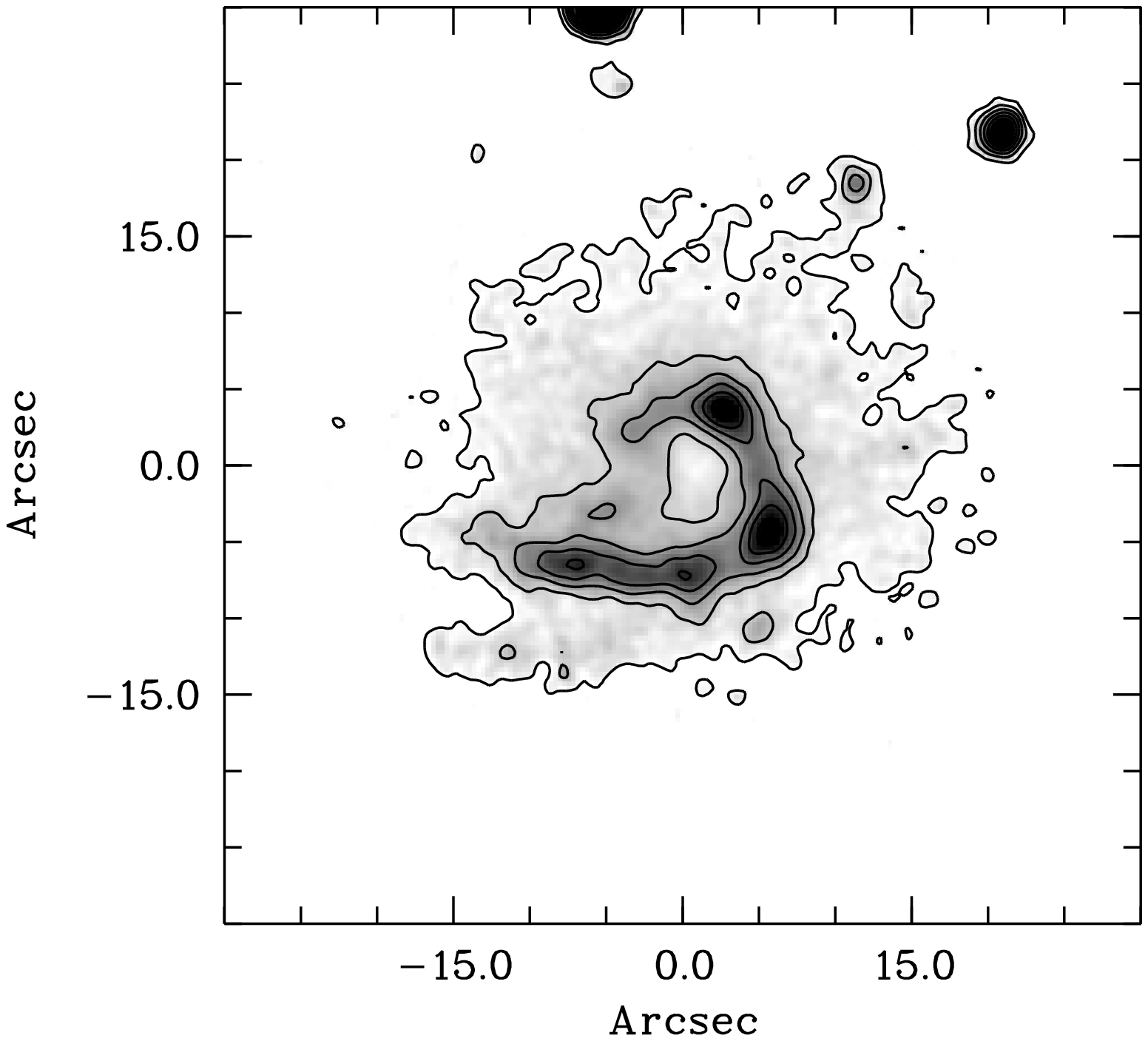,width=6.35cm}
}
\vspace*{-57.5mm}
\centerline{
\hspace*{10.77mm}
\psfig{figure=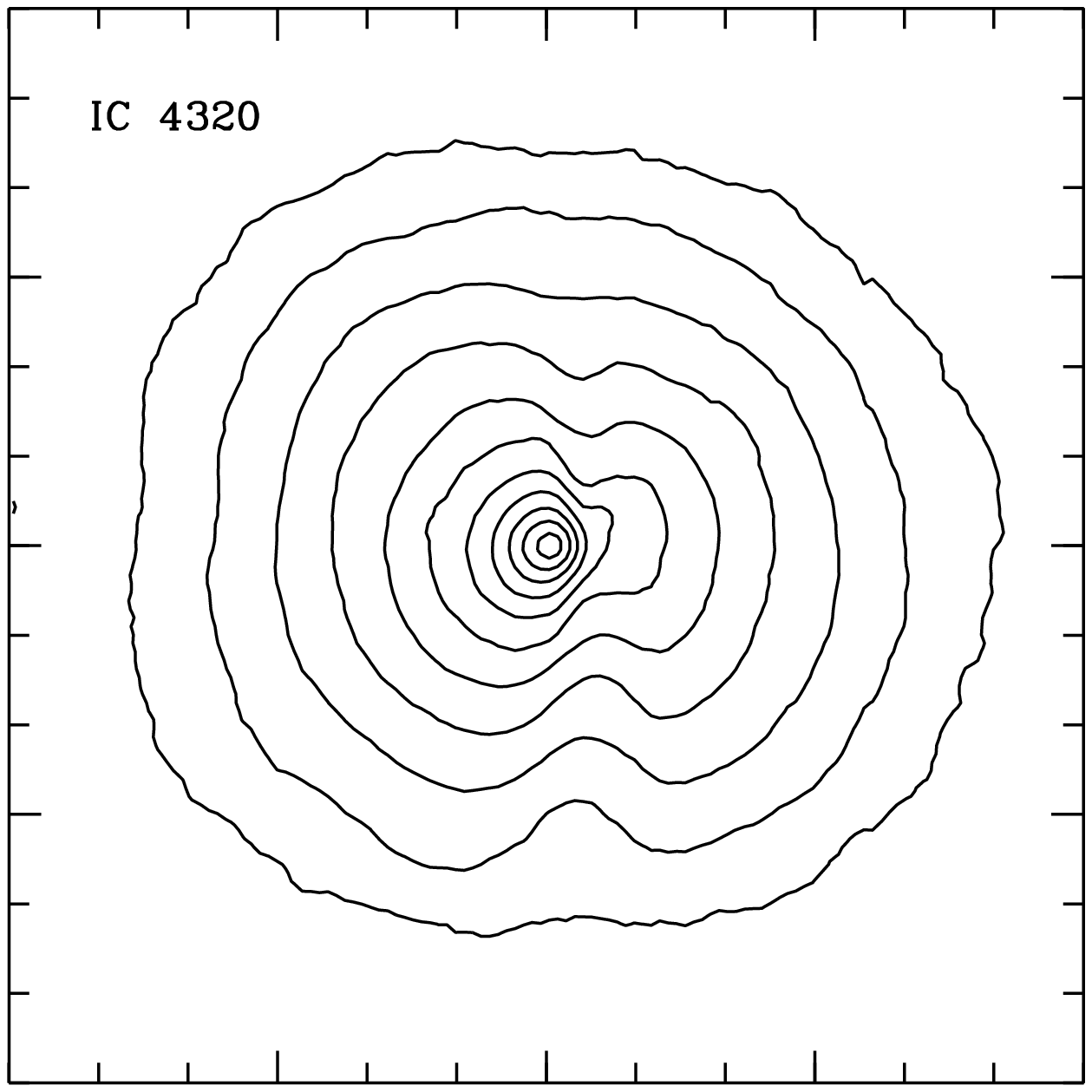,width=5.145cm,height=5.105cm}
\hfill
\psfig{figure=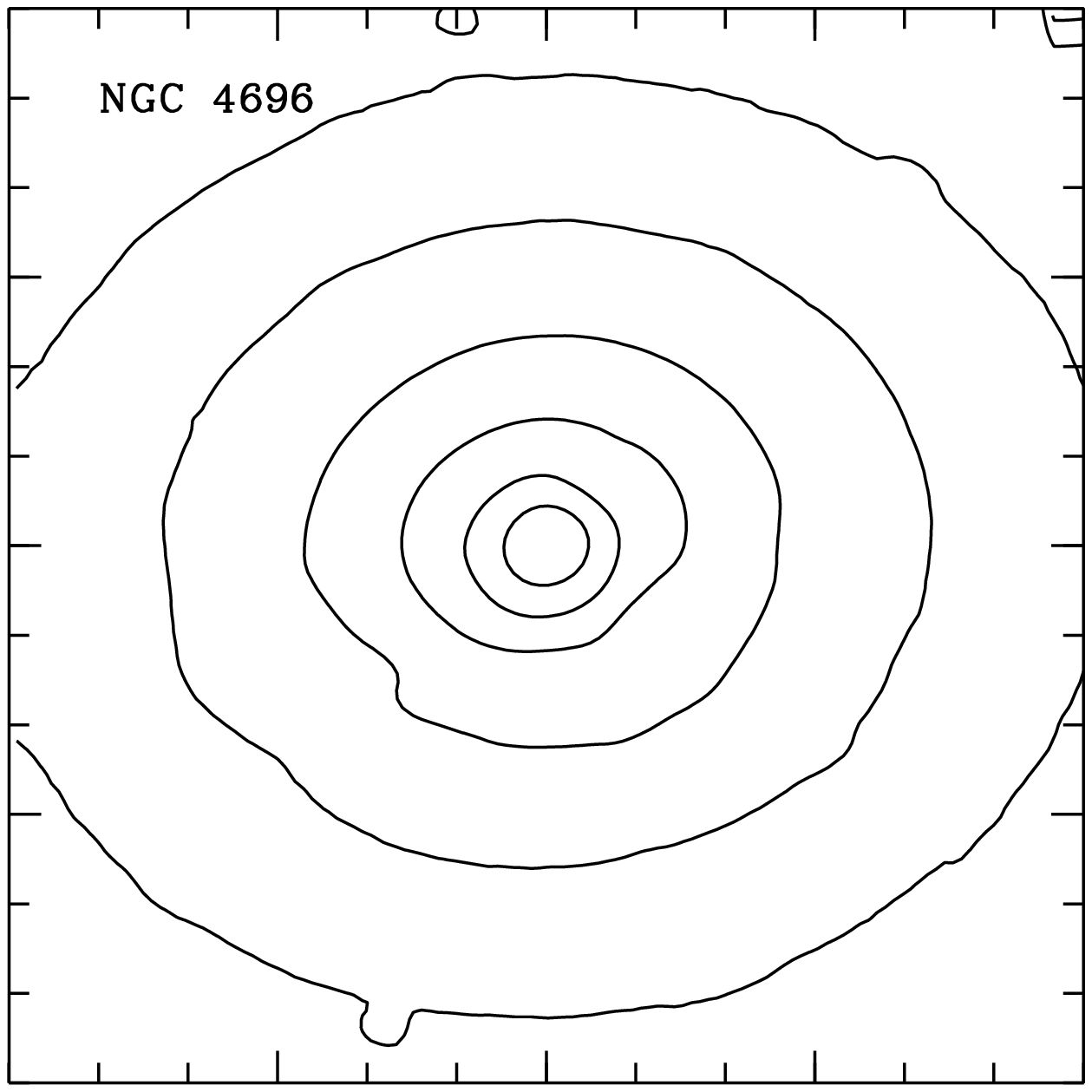,width=5.157cm,height=5.105cm}
}
\vspace*{8mm}
\centerline{
\psfig{figure=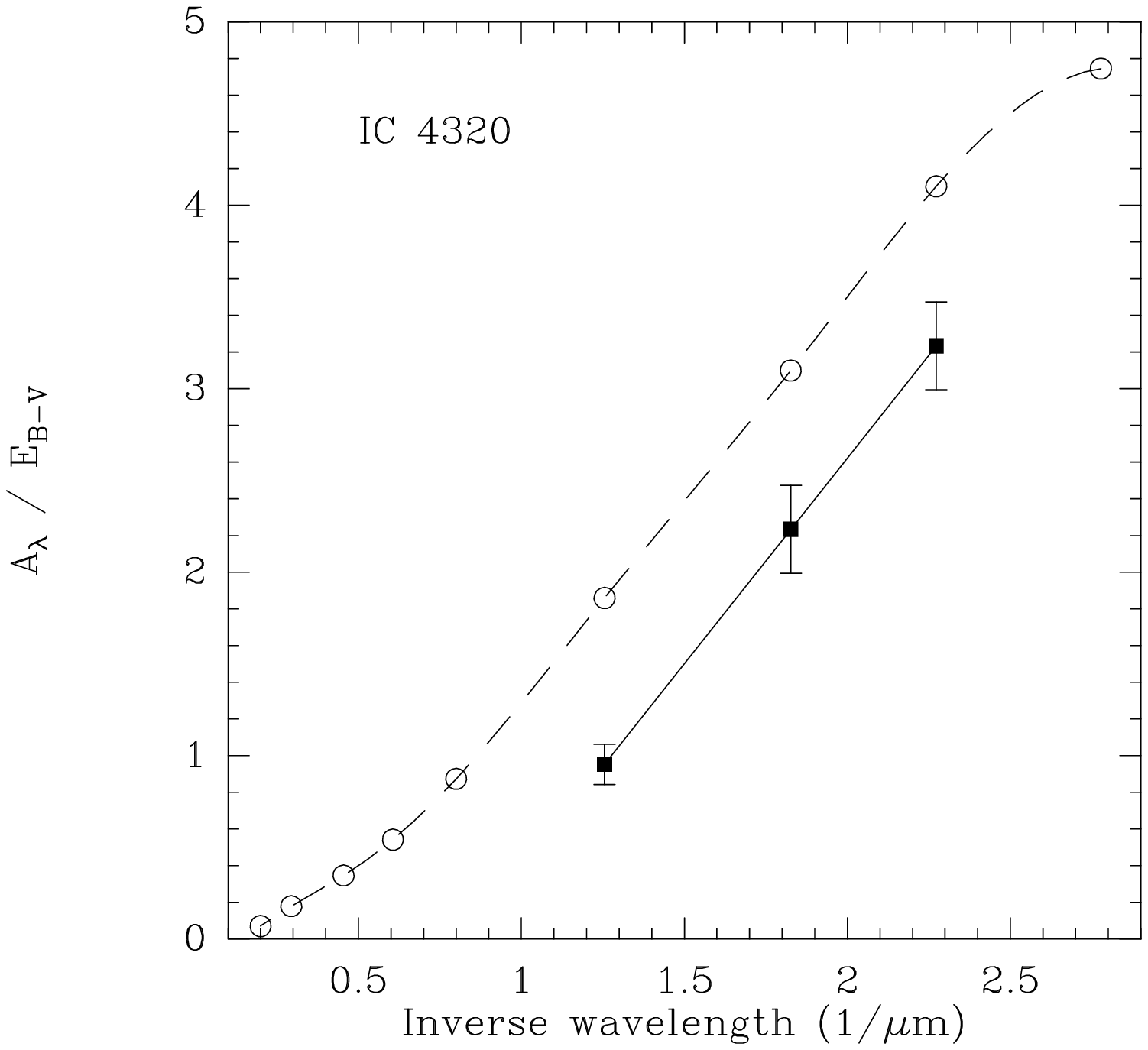,width=6.35cm}
\hfill
\psfig{figure=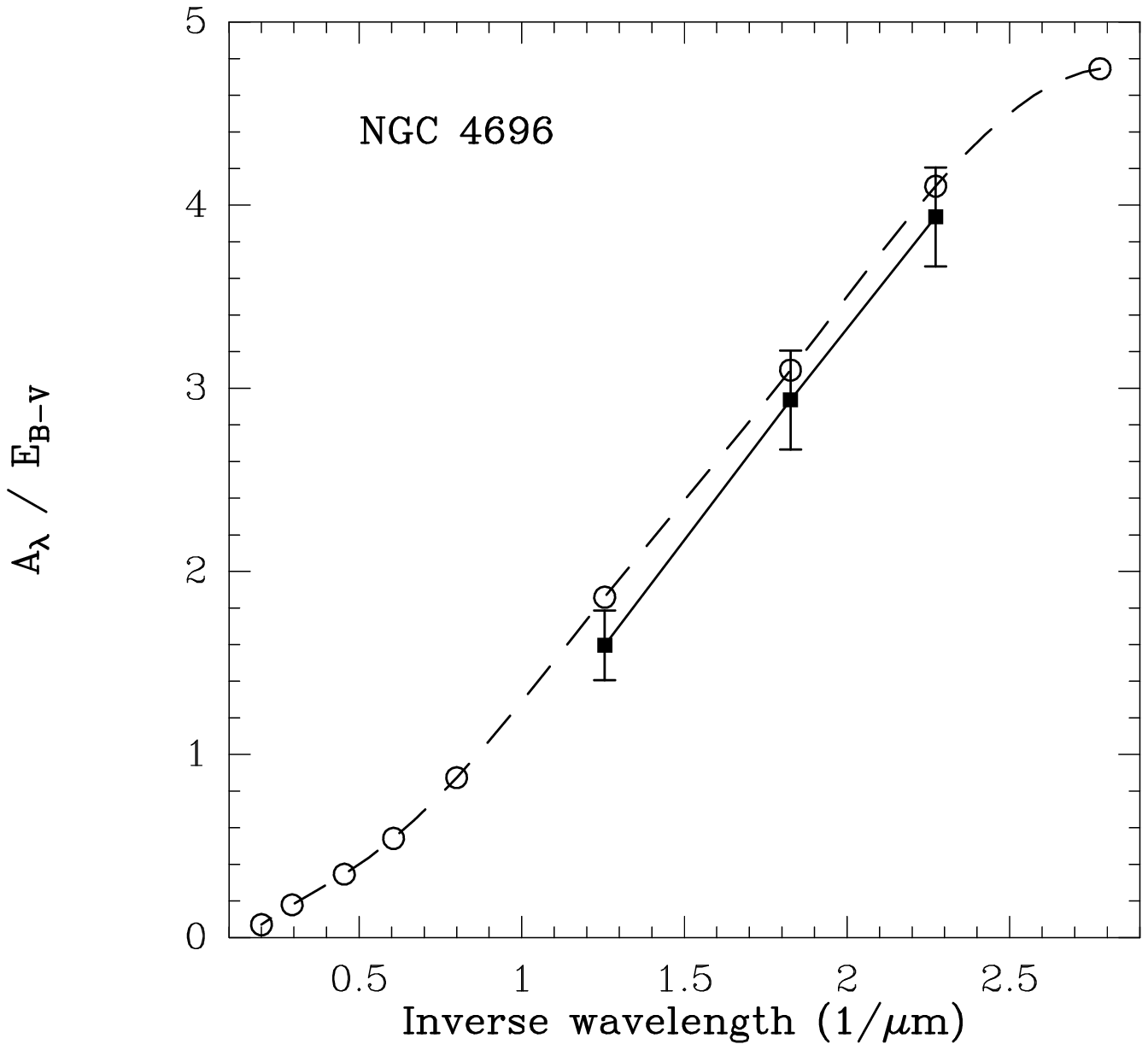,width=6.4cm}
}
\vspace*{-2mm}
\caption[]{\baselineskip=0.9\normalbaselineskip
An illustration of the difference in properties of dust in giant elliptical
galaxies in different environments (see text for details). The top
panels show $B\!-\!I$ 
color-index images (grey scales) with isophotal contours of the $B$-band
image superposed (thick solid lines). The 
bottom panels show optical ($B$,\,$V$,\,$I$) extinction curves of the galaxies
(filled squares, solid lines)
superposed on the Galactic extinction curve (open circles, dashed
lines) for comparison. {\bf Left}: IC 4320, an isolated elliptical. Note the
relaxed dust 
lane morphology and the small value of $R_V$, indicating small grains. 
{\bf Right}: NGC 4696. Note the filamentary dust morphology and the
``normal'' extinction curve. 
Images taken from Goudfrooij \etal (1994a, 1994b).
}
\label{f:fig4374_4696}
\end{figure}

Unfortunately, none of the giant ellipticals with large-scale  dust
lanes for which Goudfrooij \etal (1994b) derived $R_V$ values that are
significantly smaller than the Galactic value has been
observed in X-rays yet (except for the ROSAT all-sky survey, which was
however too shallow to detect any significant emission by hot gas from
these ellipticals). Moreover, none of them show any other sizeable 
galaxies within a radius of several 10$^2$ kpc around them on Digital Sky
Survey images, and they are not in any group 
catalog (\eg Garcia 1993). They may well be field
ellipticals which stripped dust (and gas) from small neighboring galaxies,
after which no significant replenishment of dust has occurred in the
lanes. In the absence of a massive hot gas halo, the characteristic dust
grain size will then slowly decrease through grain-grain collisions, as
observed. 

Recent models of galaxy interactions involving an elliptical-spiral pair
have shown that the settling time scale for gas disks over radial
extent of $\sim 10$ kpc (which is typical for the large-scale dust lanes in
these isolated ellipticals) is at least a few tens of crossing times
(\ie $\sim 3 \times 10^9$ yr; \eg Steiman-Cameron \& Durisen 1990). 
Note that the grain destruction time scale in hot gas ($\tau_{\rm d}
\la 10^7$ yr, 
see above) is 100\,--\,300 times shorter than this. Hence, one would predict
that ellipticals with such large-scale 
dust lanes do not host massive hot gas haloes. If this prediction proves
true, the $L_X/L_B$ ratio for these luminous ($L_B \ga 10^{11}\, \Lsun$)
ellipticals would be significantly lower than any observed so far by
{\sl EINSTEIN\/} or ROSAT. This would provide strong evidence for a scenario
in which the potential wells of single galaxies are not deep
enough to retain the stellar mass loss. Only ellipticals located in
the centers of (sub-)\-clusters or rich groups would be able to stifle
the galactic winds. One will be able to resolve this question using the
highly sensitive EPIC camera aboard the X-ray satellite {\sl XMM}.

%

\acknowledgments

It is a pleasure to thank the organizing committee of this conference
for an exciting and fruitful meeting.

\end{document}